\documentclass[fleqn,10pt]{wlscirep}

\usepackage{graphicx}
\usepackage{dcolumn}
\usepackage{bm}
\usepackage{color}
\usepackage{amsmath}
\usepackage{comment}
\usepackage{physics}

\newcommand{\fref}[1]{Fig.~\ref{#1}}

\newcommand{\eref}[1]{Eq.~(\ref{#1})}

\title{Engineering of orbital angular momentum supermodes in coupled optical waveguides}

\author[1]{A. Turpin}
\affil[1]{Departament de F\'{\i}sica, Universitat Aut\`{o}noma de Barcelona, E-08193 Bellaterra, Spain} 
\author[1,*]{G. Pelegrí}
\affil[*]{gerard.pelegri@uab.cat}
\author[1]{J.~Polo}
\author[1]{J.~Mompart}
\author[1]{V.~Ahufinger}

\begin{abstract}
In this work we demonstrate the existence of orbital angular momentum (OAM) bright and dark supermodes in a \textcolor{black}{three-evanescently} coupled cylindrical waveguides system. Bright and dark supermodes are characterized by \textcolor{black}{their} coupling and decoupling from one of the waveguides, respectively. In addition, we demonstrate that complex couplings between modes of different waveguides appear naturally due to the characteristic spiral phase-front of OAM modes in two-dimensional configurations where the waveguides are arranged forming a triangle. Finally, by adding dissipation to the waveguide uncoupled to the dark supermode, we are able to filter it out, allowing for the design of OAM mode clonners and inverters. 
\end{abstract}
\begin{document}

\flushbottom
\maketitle
\thispagestyle{empty}

\section*{Introduction}

Integrated optical devices are named to revolutionize data transfer technologies and computing platforms due to the high speed and quality of light-based communications\cite{Roadmap}. Optical waveguides are the key elements in photonic integrated circuits due to their feasible integration with additional electronic circuits. \textcolor{black}{On the one hand,} devices based on planar optical waveguides such as couplers, Mach-Zehnder interferometers, power splitters, optical modulators, wavelength demultiplexers and frequency filters have already been demonstrated \cite{book1,book2,saleh-teich,prieto:2003:nano,menchon:2013:lsa,meany:2015:lpr,obrien:2008:science}. 
\textcolor{black}{On the other hand, during the last decade there has been a significant interest in the use of multicore fibers to increase the channel capacity in optical communications through space-division multiplexing \cite{richardson:2013:natphoton,bai:2014:aop,birks:2015:aop,uden:2014:natphoton,rameez2016,mizuno2016}. In this case, the aim is to have a bundle of single mode or multimode optical fibers each of which carrying independent information integrated within a single cable. Multimode optical fibers offer the additional functionality of allowing complex structured light modes, such as light modes carrying orbital angular momentum (OAM).} 

Light beams with OAM typically possess a phase singularity in their wavefront \textcolor{black}{manifested} as a null intensity point preserved upon propagation in free-space or in cylindrically symmetric waveguides \cite{padgett:2011:aop,gabi:2007:natphys}. Although different light modes with well-defined amount of OAM have been reported, Laguerre--Gauss (LG) beams are the paradigm ones \cite{allen1992}. LG beams form a complete set of spatial modes that are \textcolor{black}{solutions} of the paraxial wave equation. They are described by Laguerre polynomials $\rm{L}_{\mathit{p}}^{|\mathit{l}|}$, where $p$ is the number of radial nodes and $l$ is the azimuthal index. In particular, they are characterized by an azimuthal term in their phase with the form $\exp(i l \phi)$, where $l$ indicates the amount of OAM carried per photon. 
Other well-known examples of light beams carrying quantized OAM per photon are Bessel beams \cite{dholakia:2002:job}, also having the characteristic azimuthal phase $\exp(i l \phi)$. Bessel beams are specially interesting since they are the fundamental family of optical modes in cylindrical waveguides \cite{book1,book2}. \textcolor{black}{At variance with polarization -only allowing for the transmission of, at most, two orthogonally polarized signals without crosstalk at a single wavelength-, OAM modes have the advantage that the dimensionality of the Hilbert space formed by OAM modes can be arbitrarily increased as one increases the number of light modes with different azimuthal indices that propagate within the same waveguide, see Refs.~\cite{padgett:2011:aop,gabi:2007:natphys,lavery:2015:aop,willner:2012:natphoton,willner:2013:science} and references therein.}
The possibility of using light beams carrying OAM adds more degrees of freedom to the control of light beams in integrated optical devices. \textcolor{black}{Since most applications of integrated optical devices take profit of the evanescent field of optical waveguides to couple two or more of these using photon tunneling, the additional degrees of freedom offered by OAM optical modes provide an alternative tool to control photon tunneling in coupled waveguides.} 
\textcolor{black}{To date, coupled waveguides carrying OAM modes have been investigated in a series of references\cite{Alexeyev2009,yavorsky:2010:jo,yavorsky:2011:jo,Li2015,yavorsky:2016:ol}.} \textcolor{black}{In Ref.~\cite{yavorsky:2011:jo}, it is shown that, in a system of two coupled waveguides, the injected light modes possessing both spin angular momentum (SAM) and OAM can tunnel to the adjacent waveguide, excite modes with opposite SAM and OAM, and that this effect could be useful for the design of couplers of optical vortices. In a similar way, in Ref.~\cite{yavorsky:2010:jo} it is investigated the propagation of higher-order modes in two weakly coupled fibers. In both cases, the spin-orbit interaction -originated by a difference between the propagation constants of two orthogonally linearly polarized supermodes- is reported to have a crucial role in the dynamics of the system.} 
\textcolor{black}{Note, however, that for weakly coupled optical waveguides with a small contrast between the indices of refraction of core and cladding, one can apply the paraxial limit and neglect the spin-orbit coupling.}

In this work, we investigate the propagation of OAM modes in a system of three cylindrical waveguides arranged in a triangular configuration. In particular, we show that photon tunneling amplitudes between OAM modes of adjacent waveguides that have opposite topological charge are, in general, complex. For the particular case of the in-line and the right triangle configurations, we demonstrate the existence of bright and dark supermodes in the system, which are characterized by \textcolor{black}{their} coupling and decoupling from the central waveguide, respectively. Thus, we discuss that any of these two configurations can be used to implement an OAM cloner and an OAM inverter by simply adding dissipation in the central waveguide while taking advantage of the projection of the input state into the dark supermode when it propagates through the system. 

\begin{figure}[]
\centering
\includegraphics[width= 0.8\columnwidth]{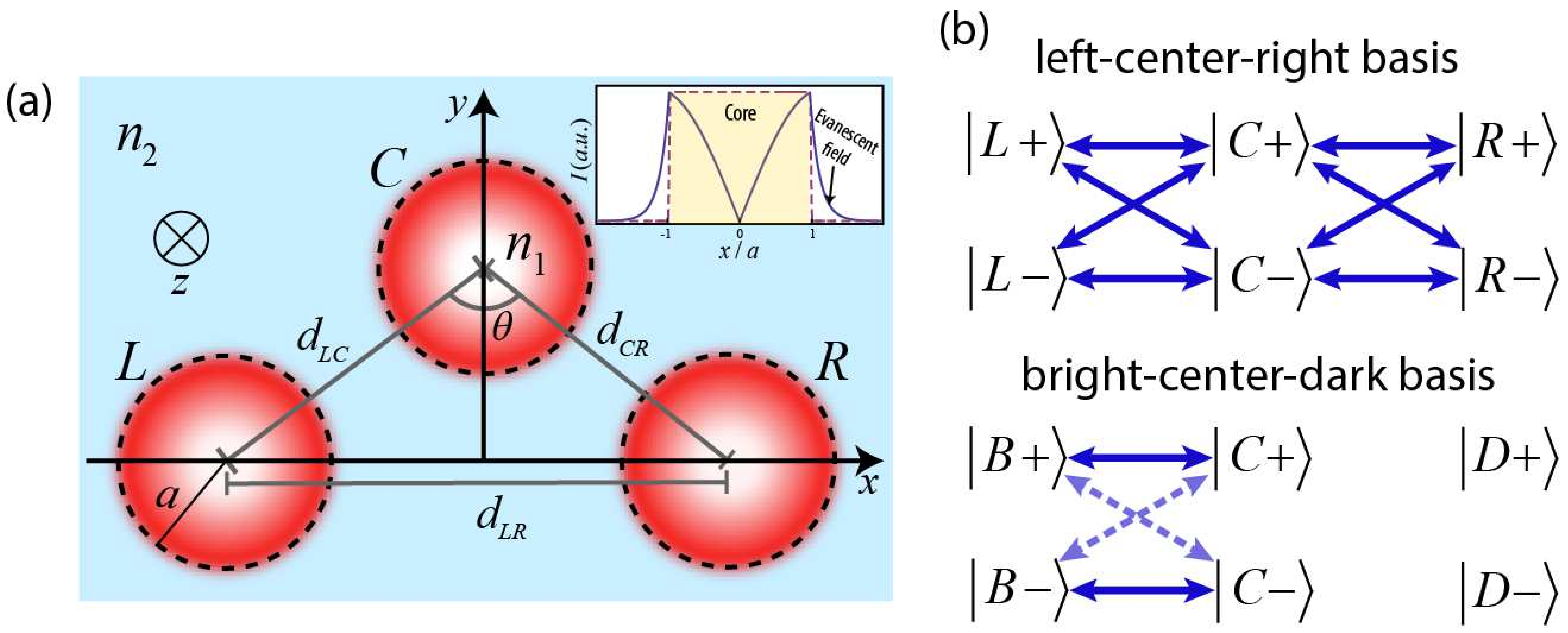}
\caption{(a) System of three identical waveguides of radius $a$ and refractive index \textcolor{black}{$n_1$} in an isosceles triangular configuration embedded in a medium of refractive index \textcolor{black}{$n_2$}. $L$, $C$, and $R$ account for the left, central and right waveguides, respectively. The inset depicts the radial intensity profile of the allowed optical modes (blue solid curve) with one unit of OAM per photon propagating along the step-index waveguides colored in yellow and delimited by the purple-dashed curve. (b) Schematic representation of the couplings between waveguides' modes in the basis left-center-right ($LCR$, top) and in the basis bright-center-dark ($BCD$, bottom). In the basis $LCR$, $\ket{L\pm}$ and $\ket{R\pm}$ modes are \textcolor{black}{only coupled to the central waveguide since left and right waveguides are assumed to be far enough to neglect its direct coupling}. \textcolor{black}{On the other hand}, in the basis \textcolor{black}{$BCD$}, bright supermodes $\ket{B\pm}$ and dark supermodes $\ket{D\pm}$ are strongly coupled and completely decoupled to the waveguide $C$, respectively.}
\label{fig1}
\end{figure}

\section*{Results}

\subsection*{OAM dark supermodes in coupled waveguides}
\label{theory}
The geometry of the optical system we consider is depicted in \fref{fig1}(a). We consider three identical \textcolor{black}{evanescently} coupled step-index cylindrical waveguides of radius $a$ and refractive index \textcolor{black}{$n_1$} embedded in a medium of refractive index \textcolor{black}{$n_2$}. We fix $d_{mn}$ as the distance between the waveguides $m$ and $n$ where $m, n = \{ L, C, R \}$, account for the left, central, and right waveguides, respectively. The three waveguides support OAM modes, and are arranged in an isosceles triangular configuration, i.e., $d_{LC} = d_{CR}(\equiv d)$. Optical modes in step-index cylindrical waveguides have the form \cite{saleh-teich}
\begin{eqnarray}
U_{ml} \left( r_m,\phi_m, z \right) &=& a_{ml}(z) u_l(r_m) e^{i l(\phi_m-\phi_0)} \textcolor{black}{e^{i \beta_{ml} z}}, \label{eq_modes1_final} \\ 
u_l(r_m) &=& 
  \begin{cases} 
   C_1 J_l(k_{\text{\tiny T}} r_m) & \text{if } r_m \leq a \\
   C_2 K_l(\alpha r_m)       & \text{if } r_m > a,
  \end{cases}
\label{eq_modes2}
\end{eqnarray}
where $\phi_0$ is a free phase parameter defining the origin of the phase, $J_l$ is the Bessel function of the first kind and order $l$, $K_l$ is the modified Bessel function of the second kind and order $l$, $k_{\text{\tiny T}}^2 = {n_1}^2 {k_0}^2 - {\beta_{ml}}^2$, \textcolor{black}{where $\beta_{ml}$ is the propagation constant} \textcolor{black}{of the mode $l$ in the waveguide $m$}, $\alpha^2 = {\beta_{ml}}^2 - {n_2}^2 {k_0}^2$, and $k_0 = \frac{2 \pi}{\lambda_0}$ is the wave-number in vacuum. $C_1$ and $C_2$ are constants satisfying \textcolor{black}{continuity at $r=a$ and} $\int |U_{ml}\left( r_m,\phi_m\right)|^2 dS_m = \mathit{P}_0$, where $P_0$ is the power of the input beam propagating through the waveguide $m$ and $dS_m = r_m d r_{m} d\phi_m$. \textcolor{black}{To simplify the formalism, we will assume (i) homogeneously linearly polarized light modes; (ii) a small contrast between the indices of refraction of core and cladding for all waveguides; and (iii) weakly evanescently coupled waveguides. In this paraxial regime, one can apply the coupled-mode equations to describe the dynamics of the OAM modes and neglect the spin-orbit coupling \cite{yavorsky:2011:jo,yavorsky:2010:jo,alexeyev:2009:jo}.}

In general, optical coupling between adjacent waveguides depends both on the form of the optical modes and on the geometry of the waveguides. For the case here considered, the coupling coefficients for two-coupled step-index waveguides $m$ and $n$ carrying $g\hbar$ and $h\hbar$ OAM per photon (with $g,h = \{+,- \}$), respectively, reads \cite{saleh-teich}:
\begin{equation}
\textcolor{black}{\kappa_{\text{\tiny mg}}^{\text{\tiny nh}} = \frac{1}{2} \left( n_1^2 - n_2^2\right) \frac{k_0^2}{\beta_{mg}} e^{i(h-g)\phi_0}
\int  u_{m}\left( r \right) e^{i g \phi_m} u_n\left( r \right) e^{-i h \phi_n} dS_n.}
\label{eq_couplings_2}
\end{equation}
\textcolor{black}{with $u_{m}(r)$ accounting for the radial part of the mode propagating in the waveguide $m$.} While for an isolated two-coupled waveguide system the origin of the phase can be set to $\phi_0 = 0$, in a system of three coupled waveguides in a triangular configuration this is only allowed for one of the two waveguides pairs. For the system under investigation, we choose the origin of the phase in the direction of $LC$. Additionally, \textcolor{black}{for all the considered configurations $d_{LR}$ is large enough compared to $d_{LC}=d_{CR}$ such that the direct coupling between the waveguides $L$ and $R$ can be neglected.} 
Light propagation in our system can be described through \textcolor{black}{coupled-mode} equations \cite{book1,book2}. These equations govern the evolution along the propagation direction of the field amplitudes, $a_{mg}$, which correspond to the $g \hbar$ OAM mode propagating through the optical waveguide $m$:
\begin{eqnarray}
i\frac{d}{dz}a_{mg}(z) = \sum\limits_{n,h} \kappa_{\text{\tiny mg}}^{\text{\tiny nh}} a_{nh}(z) \hspace{0.2cm}
\textrm{with} \hspace{0.1cm} 
m,n = L, C, R \hspace{0.2cm} \textrm{and} \hspace{0.1cm}
g,h = \pm
\label{eq_modes},
\end{eqnarray}
where $\kappa_{\text{\tiny mg}}^{\text{\tiny nh}}$ are the elements of the mode-coupling matrix ($M$) of the system. 
According to Eqs.~(\ref{eq_modes1_final})--(\ref{eq_couplings_2}) and by choosing the origin of the phase in the direction of $LC$, the coupling coefficients \textcolor{black}{satisfy the following relationships}:
\begin{gather}
\kappa_{\text{\tiny n+}}^{\text{\tiny m+}} = \kappa_{\text{\tiny n-}}^{\text{\tiny m-}} = \kappa_{\text{\tiny m+}}^{\text{\tiny n+}} = \kappa_{\text{\tiny m-}}^{\text{\tiny n-}} \equiv \kappa_{\text{\tiny 1}} \hspace{0.1cm} \text{for } n = L,R; \hspace{0.1cm} m = C, \label{k1}\\
\kappa_{\text{\tiny L+}}^{\text{\tiny C-}} = \kappa_{\text{\tiny L-}}^{\text{\tiny C+}} = \kappa_{\text{\tiny C+}}^{\text{\tiny L-}} = \kappa_{\text{\tiny C-}}^{\text{\tiny L+}} \equiv \kappa_{\text{\tiny 2}}, \label{k2} \\
{\kappa_{\text{\tiny R+}}^{\text{\tiny C-}}}^{*} = \kappa_{\text{\tiny R-}}^{\text{\tiny C+}} = {\kappa_{\text{\tiny C+}}^{\text{\tiny R-}}}^{*} = \kappa_{\text{\tiny C-}}^{\text{\tiny R+}} = \kappa_{\text{\tiny 2}} e^{i 2 \theta}\\
\textcolor{black}{\kappa_{\text{\tiny n$\pm$}}^{\text{\tiny n$\pm$}} \equiv \kappa_{\text{\tiny 0}}},
\end{gather}
where $\kappa_{\text{\tiny 0}}, \kappa_{\text{\tiny 1}},\kappa_{\text{\tiny 2}} \in \mathbb{R}$ and $\theta$ is the angle between the $LC$ and $RC$ axes, see \fref{fig1}. 
Under these conditions, \textcolor{black}{the mode-coupling matrix} $M$ becomes
\begin{equation}
M = \left( \begin{array}{cccccc}
\kappa_{\text{\tiny 0}} & 0 & \kappa_{\text{\tiny 1}} & \kappa_{\text{\tiny 2}} & 0 & 0\\
0 & \kappa_{\text{\tiny 0}} & \kappa_{\text{\tiny 2}} & \kappa_{\text{\tiny 1}} & 0 & 0\\
\kappa_{\text{\tiny 1}} & \kappa_{\text{\tiny 2}} & \kappa_{\text{\tiny 0}} & 0 & \kappa_{\text{\tiny 1}} & \kappa_{\text{\tiny 2}} e^{-i 2 \theta} \\
\kappa_{\text{\tiny 2}} & \kappa_{\text{\tiny 1}} & 0 & \kappa_{\text{\tiny 0}} & \kappa_{\text{\tiny 2}}e^{i 2 \theta} & \kappa_{\text{\tiny 1}} \\
0 & 0 & \kappa_{\text{\tiny 1}} & \kappa_{\text{\tiny 2}} e^{-i 2 \theta} & \kappa_{\text{\tiny 0}} & 0 \\
0 & 0 & \kappa_{\text{\tiny 2}}e^{i 2 \theta} & \kappa_{\text{\tiny 1}} & 0 & \kappa_{\text{\tiny 0}}
\end{array} \right),
\label{eq_hamilt_simple}
\end{equation}
expressed in the basis $\left\{  \ket{L+}, \ket{L-}, \ket{C+}, \ket{C-}, \ket{R+}, \ket{R-}\right\}$, in what follows being denoted as $LCR$ \textcolor{black}{basis}. For simplicity, we use here the standard Dirac notation of quantum mechanics, i.e., $\bra{r_m,\phi_m}\ket{m,g} \equiv U_{mg} \left( r_m, \phi_m \right)$. 
\textcolor{black}{Note that as the considered light modes of the individual waveguides have all the same spatial profile, the elements of the diagonal are all the same. As a consequence, these common elements of the diagonal can be factorized giving rise to a global phase into the dynamics.} 
\textcolor{black}{Eqs.~(\ref{k1})--(\ref{eq_hamilt_simple}) show that one can easily obtain complex coupling amplitudes between light modes of adjacent waveguides possessing opposite topological charge by simply changing the geometric arrangement of the three-coupled cylindrical waveguides.} Worth to highlight here, these complex couplings may open the way to simulate artificial gauge fields \textcolor{black}{and involved} solid-state Hamiltonians \cite{goldman:2014:review}. 

The dynamics of our system given by \eref{eq_modes} can be described in different bases. The most natural is to use the basis $LCR$, for which there is a direct correspondence between the elements of the basis and the localized OAM modes. However, one can also define a symmetric-center-antisymmetric ($SCA$) basis, where the symmetric ($S$) and antisymmetric ($A$) supermodes are defined as 
\begin{eqnarray}
\ket{S\pm} &\equiv& \frac{1}{\sqrt{2}} \left( \ket{L\pm} + \ket{R\pm} \right), \label{sym} \\
\ket{A\pm} &\equiv& \frac{1}{\sqrt{2}} \left( \ket{L\pm} - \ket{R\pm} \right). \label{antisym}
\end{eqnarray}
\textcolor{black}{Note that all along the article, we mean by supermodes those modes that extend over the full system and are not localized in a single waveguide.} For $\theta = \frac{\pi}{2}$ and $\theta = \pi$, i.e., for the right triangle configuration and for the in-line configuration, one can define the bright-central-dark \textcolor{black}{($BCD$)} basis, where the bright (B) and dark (D) supermodes read
\begin{eqnarray}
\ket{B\pm}_{\theta=\frac{\pi}{2}} &\equiv& \frac{1}{\kappa} \left( \kappa_{\text{\tiny 1}} \ket{S\pm} + \kappa_{\text{\tiny 2}} \ket{A\mp} \right) = 
\frac{1}{\sqrt{2}\kappa} \left[ \kappa_{\text{\tiny 1}} \left( \ket{L\pm} + \ket{R\pm} \right) + \kappa_{\text{\tiny 2}} \left( \ket{L\mp} - \ket{R\mp} \right) \right], \label{brights90} \\
\ket{D\pm}_{\theta=\frac{\pi}{2}} &\equiv& \frac{1}{\kappa} \left( \kappa_{\text{\tiny 2}} \ket{S\pm} - \kappa_{\text{\tiny 1}} \ket{A\mp} \right) =
\frac{1}{\sqrt{2}\kappa} \left[ \kappa_{\text{\tiny 2}} \left( \ket{L\pm} + \ket{R\pm} \right) - \kappa_{\text{\tiny 1}} \left( \ket{L\mp} - \ket{R\mp} \right) \right], \label{darks90} \\
\ket{B\pm}_{\theta=\pi} &\equiv& \frac{1}{\sqrt{2}} \left( \ket{S+} \pm \ket{S-} \right) = \frac{1}{2} \left( \ket{L+} \pm \ket{L-} + \ket{R+} \pm \ket{R-} \right), \label{brights0} \\
\ket{D\pm}_{\theta=\pi} &\equiv& \frac{1}{\sqrt{2}} \left( \ket{A+} \pm \ket{A-} \right) = \frac{1}{2} \left( \ket{L+} \pm \ket{L-} - \ket{R+} \mp \ket{R-} \right), \label{darks0} 
\end{eqnarray}
with $\kappa \equiv \sqrt{{\kappa_{\text{\tiny 1}}}^2 + {\kappa_{\text{\tiny 2}}}^2}$. 
\textcolor{black}{For $\theta = \pi /2$, $\{\ket{B+},\ket{C+},\ket{D+}\}$ and  $\{\ket{B-},\ket{C-},\ket{D-}\}$ are two uncoupled subspaces of the full Hilbert space,  while for $\theta = \pi $ one can define the two following uncoupled subspaces $\{\ket{B+},\ket{C_S},\ket{D+}\}$ and $\{\ket{B-},\ket{C_A},\ket{D-}\}$ with $\ket{C_S}=\frac{1}{\sqrt{2}}(\ket{C+}+\ket{C-})$ and $\ket{C_A}=\frac{1}{\sqrt{2}}(\ket{C+}-\ket{C-})$. In both cases, $\braket{C\pm}{D\pm} = \braket{B\pm}{D\pm} = 0$ and $\braket{B\pm}{C\pm} \neq 0$.} 
Therefore, the bright (dark) supermodes are characterized by \textcolor{black}{their} coupling (decoupling) with the central waveguide, as sketched in \fref{fig1}(b), in a similar way that bright (dark) states are coupled (uncoupled) to an intermediate state in three-level atomic systems in quantum optics \cite{orriols:1979:lnc,orriols:1979:nc}. \textcolor{black}{Note that bright supermodes are not eigenmodes of the full system while dark supermodes are.} In what follows, we will use the $BCD$ basis to understand the dynamics of OAM modes propagating in the three coupled waveguides system.

\subsection*{Dynamics of OAM supermodes in the presence of dissipation}
\label{in-line}

\begin{figure*}[]
\centering
\includegraphics[width= 1 \columnwidth]{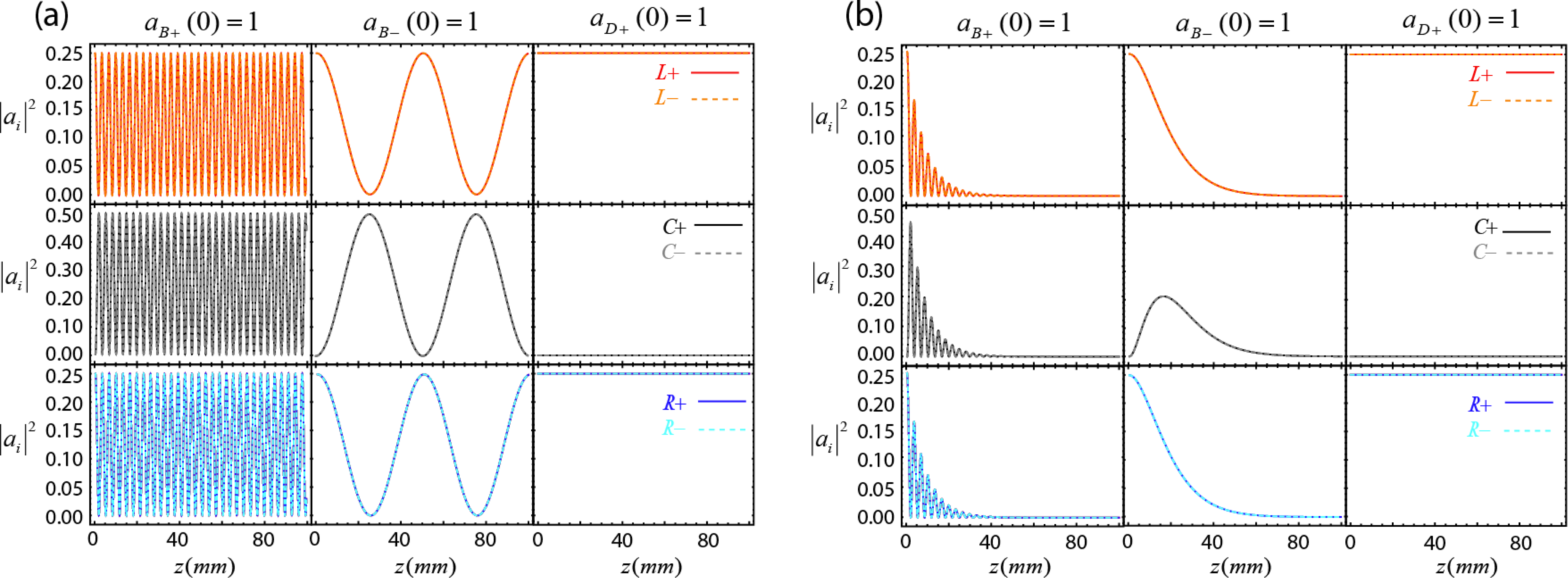}
\caption{Intensity evolution along the $z$ direction for the $\ket{L+}$ (red-solid curve), $\ket{L-}$ (orange-dashed curve), $\ket{C+}$ (black-solid curve), $\ket{C-}$ (gray-dashed curve), $\ket{R+}$ (blue-solid curve), and $\ket{R-}$ (turquoise-dashed curve) OAM modes when $\ket{B+}$ (first column), $\ket{B-}$ (second column), and $\ket{D+}$ (third column) are injected to the system. (a): Without absorption along the waveguide $C$. (b): With a power absorption coefficient \textcolor{black}{$\gamma = 250\,\rm{m^{-1}}$} for the waveguide $C$. Parameters used in the simulations: $\lambda = 1.55\,\rm{\mu m}$, \textcolor{black}{$n_1 = 1.52$, $n_2 = 1.51$,} $a = 10\,\rm{\mu m}$, $\textcolor{black}{d = 2.4a}$, \textcolor{black}{$\kappa_{\text{\tiny 1}} = 362\,\rm{m^{-1}}$, and $\kappa_{\text{\tiny 2}} = 318\,\rm{m^{-1}}$}.}
\label{fig2}
\end{figure*}

We focus on the case $\theta = \pi$, i.e., the in-line configuration. However, note that since our approach is based on the use of bright and dark supermodes, the forthcoming results are also found in the right triangular configuration.  Fig.~\ref{fig2}(a) shows plots of the evolution along the $z$ axis of the OAM modes intensities according to our six-state model when injecting a bright and a dark supermode given by Eqs.~(\ref{brights0}) and (\ref{darks0}) into the three-waveguide system, i.e., for $a_{B+}(0) = 1$ (left column), $a_{B-}(0) = 1$ (central column), and $a_{D+}(0) = 1$ (right column; analogous results are obtained for $a_{D-}(0) = 1$). The first row corresponds to the intensity evolution of $\ket{L+}$ ($|a_{L+}|^2$, red solid line) and $\ket{L-}$ ($|a_{L-}|^2$, orange dashed line); the second row corresponds to $\ket{C+}$ ($|a_{C+}|^2$, black solid line) and $\ket{C-}$ ($|a_{C-}|^2$, gray dashed line); and the third row corresponds to $\ket{R+}$ ($|a_{R+}|^2$, blue solid line) and $\ket{R-}$ ($|a_{R-}|^2$, turquoise dashed line) OAM modes. Bright supermodes are strongly coupled to the central waveguide, in particular the supermode $\ket{B+}$, as it can be clearly seen from the fast intensity oscillations. In contrast, dark supermodes are completely decoupled from the central waveguide. \textcolor{black}{Note also that the fast (slow) spatial oscillation frequency when the input mode is the bright supermode $\ket{B+}$ ($\ket{B-}$) is given by the coupling constant between states $\ket{B+}$ and $\ket{C_S}$ ($\ket{B-}$ and $\ket{C_A}$), which for the in-line configuration is $\sqrt{2}(\kappa_1+\kappa_2 )$ ($\sqrt{2}(|\kappa_1-\kappa_2|)$), see Eqs.~(\ref{coup34}-\ref{coup37}) in the Methods section.}  \textcolor{black}{We also note that, for the separation between the center of the waveguides that we consider here, $d = 2.4a$, we have obtained that $\kappa_{\text{\tiny 0}} = 40413\,\rm{m^{-1}}$, which is much larger than $\kappa_{\text{\tiny 1}} = 362\,\rm{m^{-1}}$ and $\kappa_{\text{\tiny 2}} = 318\,\rm{m^{-1}}$.Thus, we are well within the weak coupling regime for which the couple-mode equations are valid.}

Now, we will consider that the central waveguide absorbs light by replacing the corresponding two equations for the $a_{C+}$ and $a_{C-}$ \textcolor{black}{amplitudes} in \eref{eq_modes} by: 
\begin{equation}
i\frac{d}{dz}a_{Cg}(z) = \textcolor{black}{\sum\limits_{n,h} \kappa_{Cg}^{nh} a_{nh}(z)} - i \frac{\gamma}{2} a_{Cg}(z),\label{eq_modes_dissipation}
\end{equation}
\textcolor{black}{where $n = L,C,R$, $g,h = \pm$, and $\gamma$ is the power absorption coefficient.}
\fref{fig2}(b) shows plots of the evolution along the $z$ axis of the OAM modes intensities by numerical integration of the coupled-mode equations of the six-state model when injecting bright and dark supermodes into the three-waveguide system with the absorption coefficient \textcolor{black}{$\gamma = 250\,\rm{m^{-1}}$} in the waveguide $C$. As it can be appreciated, the presence of absorption makes the dynamics of the system particularly interesting since the supermodes $\ket{B\pm}$ are completely absorbed after a certain propagation distance and only the supermodes $\ket{D\pm}$ are transmitted loseless through the system.

\section*{Discussion}

\begin{figure*}[]
\centering
\includegraphics[width= 0.7 \columnwidth]{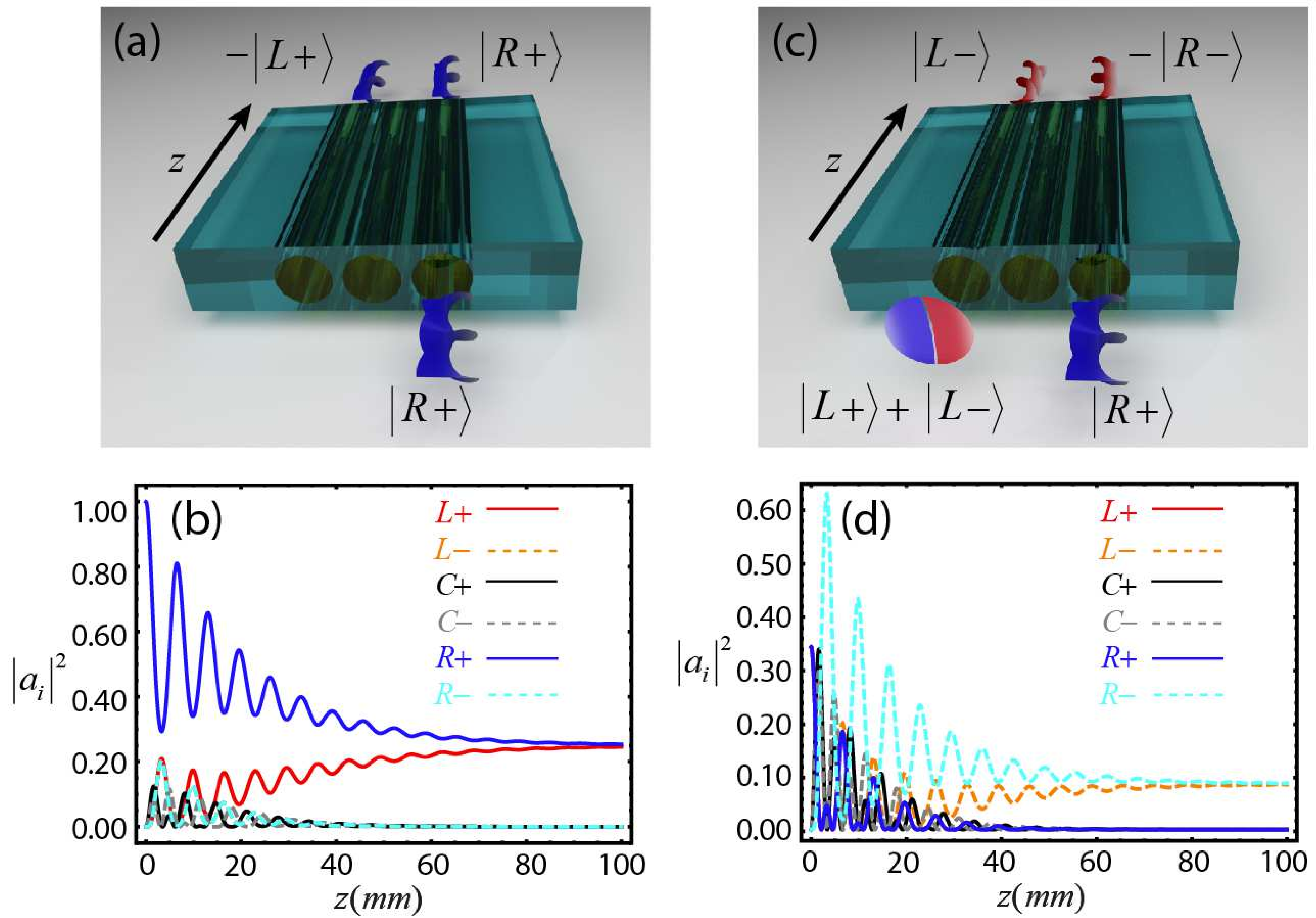}
\caption{\textcolor{black}{Examples of OAM modes engineering in an in-line three-coupled waveguide system assisted by dissipation in the central waveguide. (a) Illustration of an OAM cloner of the $\ket{R+}$ mode to the waveguide $L$, and (b) the corresponding intensity evolution along the $z$ direction of $\ket{L+}$ (red-solid curve), $\ket{L-}$ (orange-dashed curve), $\ket{C+}$ (black-solid curve), $\ket{C-}$ (gray-dashed curve), $\ket{R+}$ (blue-solid curve), and $\ket{R-}$ (turquoise-dashed curve) OAM modes. (c) Illustration of an OAM switcher of the $\ket{R+}$ mode to the $\ket{R-}$ mode and (d) the corresponding intensity evolution along the $z$ direction.
Same parameters as in Fig.~\ref{fig2} were used, together with an absorption coefficient$\gamma = 250\,\rm{m^{-1}}$ at the waveguide $C$.}}
\label{fig3}
\end{figure*}

To show the suitability of dissipation to control the dynamics of OAM supermodes in coupled waveguides systems, in what follows we show how to implement a mode cloner and a mode inverter in an in-line configuration ($\theta = \pi$) with absorption at the central waveguide.  
The absorption of supermodes $\ket{B\pm}$ can be used to engineer 
the outcoming OAM modes at the waveguides $L$ and $R$. 
For instance, from Eqs.~(\ref{brights0}) and (\ref{darks0}), one can write that $ 2\ket{R+} = \ket{B+} + \ket{B-} - \ket{D+} - \ket{D-}$. 
Since we have shown that $\ket{B\pm}$ supermodes are absorbed by the waveguide $C$ and that $\ket{D\pm}$ supermodes are completely decoupled from it, by injecting $\ket{R+}$ the output beam after a certain propagation distance becomes \textcolor{black}{$-(\ket{D+} + \ket{D-}) = 2 \left( \ket{R+} - \ket{L+} \right)$}. Therefore, one obtains with a $50\%$ efficiency the superposition state $\ket{R+}-\ket{L+}$, which means that the input state can be cloned at waveguide $L$, as sketched in \fref{fig3}(a). The device here discussed can also be envisaged as a robust coherent beam spliter with $50\%$ efficiency. 
Fig.~\ref{fig3}(b) shows plots of the axial evolution of the OAM modes intensities in an OAM cloning process when $\ket{R+}$ is injected. As it can be observed, the intensities of the output modes emerging from waveguides $R$ and $L$ are identical after some propagation distance. Similar results are obtained for other input states, i.e., for an input state of the form $\ket{m\pm}$, the outcome $\ket{n\pm} - \ket{m\pm}$ is expected, where $m,n = \{L,R\}$ and $m \neq n$. Note that it is not possible to \textcolor{black}{efficiently} split an OAM mode between two-evanescent parallel coupled waveguides. \textcolor{black}{When two waveguides are directly coupled, the input OAM mode in one of the fibers couples with the  OAM modes of the adjacent fiber that possess the same but also opposite topological charge.} \textcolor{black}{Thus, the dynamics becomes very involved and at all propagation distances the state of the system is a superposition of all OAM modes with positive and negative topological charges.} 

Furthermore, the coupling between $\ket{B\pm}$ supermodes with the absorbing central waveguide leads to a counterintuitive result: states $\ket{R\pm} + \ket{L\pm}$ are completely dissipated when propagating through the system, since they have null projection with the dark supermodes. As a consequence, input states with the form $\ket{m\pm} + \ket{m\mp} + \ket{n\pm}$, which have $0 \hbar$ OAM at the waveguide $m$ and $\pm \hbar$ OAM at the waveguide $n$, emerge as $\ket{m\mp} - \ket{n\mp}$ from the system, where $m,n = \{L,R\}$ with $m \neq n$. Thus, one can induce net OAM at the waveguide $m$ and control its sign by appropriately choosing the input state at the waveguide $n$, and vice versa. An alternative interpretation is that this configuration can be used to invert the sign of the OAM state at the waveguide $n$ by injecting a state with null OAM at the waveguide $m$. To better visualize this concept, in \fref{fig3}(c) we show the case $\ket{L+} + \ket{L-} + \ket{R+} \rightarrow \ket{L-} - \ket{R-}$, where the blue and red spirals indicate light modes carrying $\pm \hbar$ OAM per photon, respectively. In Fig.~\ref{fig3}(d) we plot the evolution along the $z$ axis of the OAM mode intensities when $\ket{L+} + \ket{L-} + \ket{R+}$ is injected (\textcolor{black}{$\gamma = 250\,\rm{m^{-1}}$} at the waveguide $C$). As it can be observed, after some propagation distance, only the dark supermode contribution of the input state, proportional to $\ket{D+}+\ket{D-}= \ket{L-}-\ket{R-}$ survives, which corresponds to a coherent superposition of two modes propagating in waveguides $L$ and $R$ with $-\hbar$ OAM each. 

We have further investigated the role of the dissipation on the dynamics of the propagating modes. Our calculations reveal that for high values of the absorption coefficient at the waveguide $C$, i.e., for \textcolor{black}{$\gamma \gg \kappa_{\text{\tiny 1}},\kappa_{\text{\tiny 2}}$}, the waveguides $R$ and $L$ become completely decoupled from waveguide $C$. This scenario resembles the quantum Zeno effect associated to the dynamical inhibition of the population excitation for a coherently driven atomic system under continuous observation of its fluorescence \cite{zeno:1977,zeno:2000}. From numerical simulations with the parameters of the system here investigated, we have checked that Zeno-like regimes only appear for \textcolor{black}{$\gamma > 10^4 \kappa_{\text{\tiny 1}}$}.

\section*{Conclusions}
In summary, in this work we have demonstrated the existence of OAM bright and dark optical supermodes in three-evanescently coupled step-index cylindrical waveguides. Bright and dark supermodes are characterized by \textcolor{black}{their} coupling and decoupling from the central waveguide of the system, respectively. Under this scenario, we have shown that the output optical modes from the waveguides can be engineered by adding dissipation to the central waveguide, which makes the system  absorb the bright supermodes when they propagate through the system. In particular, we have proposed the use of the bright and dark OAM supermodes for cloning of the input OAM mode into another waveguide and also for inverting, i.e. changing, the sign of the OAM mode propagating along one of the waveguides. In addition to the possibility of controlling OAM modes in coupled waveguides, we recall the interest of our approach due to the natural appearance of complex couplings depending on the angle between waveguides in a triangular configuration. This method introduces a new degree of freedom to control phases in photonic quantum simulators \cite{longhi:2009:lpr,walther:2012:natphys}. \textcolor{black}{Although we have restricted our analysis to a basic geometry based on three-coupled waveguides, the here shown results could be applied to waveguides arrays having in-line or right triangle three-coupled waveguides as unit cells to provide novel alternatives in the field of space-division multiplexing with multicore fibers.} \textcolor{black}{Finally, it would be also interesting to extend our previous study to physical scenarios where the spin-orbit coupling plays a significant role and find out whether bright and dark states are also present there.}


\section*{Methods}

\subsection*{Coupling coefficients in the symmetric-center-antisymmetric basis}
The $SCA$ basis formed by symmetric ($S$), central ($C$), and antisymmetric ($A$) supermodes, is defined by states $\ket{C\pm}$ and:
\begin{eqnarray}
\ket{S\pm} &\equiv& \frac{1}{\sqrt{2}} \left( \ket{L\pm} + \ket{R\pm} \right), \label{ssym} \\
\ket{A\pm} &\equiv& \frac{1}{\sqrt{2}} \left( \ket{L\pm} - \ket{R\pm} \right). \label{aantisym}
\end{eqnarray}
By taking into account the form of the OAM modes allowed in step-index cylindrical waveguides, it is straightforward to obtain the coupling coefficients between $\ket{S\pm}$, $\ket{A\pm}$, and $\ket{C\pm}$: 
\begin{eqnarray}
\kappa_{\text{\tiny qg}}^{\text{\tiny th}} &=& 0 \hspace{0.2cm} \text{for } q,t = \{S,A\}; \hspace{0.1cm} g,h = \pm, \label{nullcoup}\\
\kappa_{\text{\tiny S$\pm$}}^{\text{\tiny C$\pm$}} &=&  \sqrt{2} \kappa_{\text{\tiny 1}}, \label{nullcoup2}\\
\kappa_{\text{\tiny S$\pm$}}^{\text{\tiny C$\mp$}} &=&  \frac{1}{\sqrt{2}} \kappa_{\text{\tiny 2}}\left(1 + e^{\pm i 2 \theta} \right), \label{ksc} \\ 
\kappa_{\text{\tiny A$\pm$}}^{\text{\tiny C$\pm$}} &=&  0,\\
\kappa_{\text{\tiny A$\pm$}}^{\text{\tiny C$\mp$}} &=&  \frac{1}{\sqrt{2}} \kappa_{\text{\tiny 2}}\left(1 - e^{\pm i 2 \theta} \right), \label{kac}
\end{eqnarray}
where $\kappa_{\text{\tiny 1}}$ and $\kappa_{\text{\tiny 2}}$ are given by Eqs.~(\ref{k1}) and (\ref{k2}), respectively.

\subsection*{Coupling coefficients in the bright-center-dark basis}
We define two $BCD$ bases from the symmetric and antisymmetric supermodes defined above for the a) right triangular and b) in-line configurations. 
\subsubsection*{a) Right triangle configuration.}
By defining 
\begin{eqnarray}
\ket{F\pm} &\equiv& \frac{1}{\kappa} \left( \kappa_{\text{\tiny 1}} \ket{S\pm} + \kappa_{\text{\tiny 2}} \ket{A\mp} \right), \label{bpm1} \\
\ket{G\pm} &\equiv& \frac{1}{\kappa} \left( \kappa_{\text{\tiny 1}} \ket{S\pm} - \kappa_{\text{\tiny 2}} \ket{A\mp} \right), \label{dpm1}
\end{eqnarray}
where $\kappa = \sqrt{\kappa_{\text{\tiny 1}}^2+\kappa_{\text{\tiny 2}}^2}$, the coupling coefficients between $\ket{F\pm}$, $\ket{G\pm}$, and $\ket{C\pm}$ can be obtained from Eqs.~(\ref{nullcoup})--(\ref{kac}): 
\begin{eqnarray}
\kappa_{\text{\tiny qg}}^{\text{\tiny th}} &=& 0 \hspace{0.2cm} \text{for } q,t = \{F,G\}; \hspace{0.1cm} g,h = \pm, \label{nullcoup2}\\
\kappa_{\text{\tiny F$\pm$}}^{\text{\tiny C$\pm$}} &=&  \frac{1}{\kappa} \left[ \sqrt{2} \kappa_{\text{\tiny 1}}^2 + \frac{\kappa_{\text{\tiny 2}}^2}{\sqrt{2}} \left( 1 - e^{\pm i 2 \theta} \right) \right], \label{coupb1cpm}\\
\kappa_{\text{\tiny F$\pm$}}^{\text{\tiny C$\mp$}} &=&  \frac{\kappa_{\text{\tiny 1}}\kappa_{\text{\tiny 2}}}{\sqrt{2} \kappa} \left(1 + e^{\mp i 2 \theta} \right), \label{coupb1cmp} \\ 
\kappa_{\text{\tiny G$\pm$}}^{\text{\tiny C$\pm$}} &=&  \frac{\kappa_{\text{\tiny 1}} \kappa_{\text{\tiny 2}}}{\sqrt{2}\kappa} \left[ 1 + e^{\pm i 2 \theta} \right], \label{coupd1cpm}\\
\kappa_{\text{\tiny G$\pm$}}^{\text{\tiny C$\mp$}} &=&  \frac{\kappa_{\text{\tiny 1}}\kappa_{\text{\tiny 2}}}{\sqrt{2} \kappa} \left(1 + e^{\mp i 2 \theta} \right) \label{coupd1cmp}.
\end{eqnarray}
Note that, for $\theta = \frac{\pi}{2}$, $\kappa_{\text{\tiny G$\pm$}}^{\text{\tiny C$\pm$}} = \kappa_{\text{\tiny G$\pm$}}^{\text{\tiny C$\mp$}} =0$, which means that supermodes $\ket{G\pm}$ become completely decoupled from waveguide $C$ and we recover expressions (\ref{darks90}) for the dark supermodes and (\ref{brights90}) for the bright ones. 

\subsubsection*{b) In-line configuration.}
Analogously, we define:
\begin{eqnarray}
\ket{\tilde{F}\pm} &\equiv& \frac{1}{\sqrt{2}} \left( \ket{S+} \pm \ket{S-} \right), \label{bpm2} \\
\ket{\tilde{G}\pm} &\equiv& \frac{1}{\sqrt{2}} \left( \ket{A+} \pm \ket{A-} \right), \label{dpm2}
\end{eqnarray}
with couplings \textcolor{black}{
\begin{eqnarray}
\kappa_{\text{\tiny qg}}^{\text{\tiny t h}} &=& 0 \hspace{0.2cm} \text{for } q,t = \{B_2,D_2\}; \hspace{0.1cm} g,h = \pm, \label{nullcoup3}\\
 \kappa_{\text{\tiny $\tilde{F}$$+$}}^{\text{\tiny C$\pm$}}  &=&  \frac{\kappa_{\text{\tiny 2}}}{2}\left( 1 + e^{\mp i 2 \theta} \right)  + \kappa_{\text{\tiny 1}},
 \label{coup34}\\
 \kappa_{\text{\tiny $\tilde{F}$$-$}}^{\text{\tiny C$\pm$}} &=& \mp \left(  \frac{\kappa_{\text{\tiny 2}}}{2}\left( 1 + e^{\mp i 2 \theta} \right)  - \kappa_{\text{\tiny 1}} \right),
 \label{coup35}\\
 \kappa_{\text{\tiny $\tilde{G}$$+$}}^{\text{\tiny C$\pm$}}    &=&  \frac{\kappa_{\text{\tiny 2}}}{2} \left( 1 - e^{\mp i 2 \theta} \right),
\label{coup36}\\
 \kappa_{\text{\tiny $\tilde{G}$$-$}}^{\text{\tiny C$\pm$}}    &=& \mp  \frac{\kappa_{\text{\tiny 2}}}{2} \left( 1 - e^{\mp i 2 \theta} \right),
 \label{coup37}
\end{eqnarray}
}
In this case, decoupling between the waveguide $C$ and supermodes $\tilde{G}\pm$ is obtained for $\theta = \pi$ recovering expressions (\ref{brights0}) and (\ref{darks0}) for the bright and dark supermodes, respectively. 

\subsection*{Parameters used in the numerical simulations of the coupled-mode equations}
All numerical simulations have been carried out using the following parameters: wavelength $\lambda = 1.55\,\rm{\mu m}$, \textcolor{black}{$n_1 = 1.52$, $n_2 = 1.51$,} radius of the waveguides $a = 10\,\rm{\mu m}$, distances between waveguides \textcolor{black}{$d_{LR} = d_{RC} \equiv d = 2.4a$}. \textcolor{black}{With these parameters, the coupling coefficients $\kappa_1$ and $\kappa_2$ given by Eqs.~(\ref{k1})--(\ref{k2}) are: $\kappa_{\text{\tiny 1}} = 362\,\rm{m^{-1}}$, and $\kappa_{\text{\tiny 2}} = 318\,\rm{m^{-1}}$. The absorption coefficient of the central waveguide is $\gamma = 250\,\rm{m}^{-1}$.}
All numerical simulations have been carried out using the following parameters: wavelength $\lambda = 1.55\,\rm{\mu m}$, \textcolor{black}{$n_1 = 1.52$, $n_2 = 1.51$,} radius of the waveguides $a = 10\,\rm{\mu m}$, distances between waveguides \textcolor{black}{$d_{LR} = d_{RC} \equiv d = 2.4a$}. \textcolor{black}{With these parameters, the coupling coefficients $\kappa_1$ and $\kappa_2$ given by Eqs.~(\ref{k1})--(\ref{k2}) are: $\kappa_{\text{\tiny 1}} = 362\,\rm{m^{-1}}$, and $\kappa_{\text{\tiny 2}} = 318\,\rm{m^{-1}}$. The absorption coefficient of the central waveguide is $\gamma = 250\,\rm{m}^{-1}$.}

\section*{Acknowledgements}

The authors gratefully acknowledge financial support through the Spanish Ministry of Science and Innovation (MINECO) (Contract No. FIS2014-57460P) and the Catalan Government (Contract No. SGR2014-1639). J.P. acknowledges financial support from the MICINN through the Grant No. BES-2012-053447. G.P. acknowledges financial support from the MICINN through the Grant No. BES-2015-073772.  

\section*{Author contributions statement}

J. P. and G. P. developed the six-state model. A. T. made the numerical simulations and conceived the idea of using dissipation to engineer OAM modes. J. M. and V. A. proposed and supervised the project. All authors wrote the manuscript.

\section*{Additional information}

\textbf{Competing financial interests}

\noindent The authors declare no competing financial interests.

\end{document}